\documentclass[%
 amsmath,amssymb,
 aps,
 prb,twocolumn
]{revtex4-1}

\usepackage{graphicx}
\usepackage{color}
\usepackage{bm}

\newcommand{\angstrom}{\textup{\AA}}

\begin{document}

\title{Detection of topological phase transitions through entropy measurements:\\
The case of germanene
}

\author{D. Grassano}
\affiliation{Deptartment of Physics, INFN, University of Rome Tor Vergata, Via della Ricerca Scientifica 1, I-00133 Rome, Italy}
\email{davide.grassano@roma2.infn.it}

\author{O. Pulci}
\affiliation{Dept. of Physics, and INFN, University of Rome Tor Vergata, Via della Ricerca Scientifica 1, I-00133 Rome, Italy}

\author{V.O. Shubnyi}
\affiliation{Department of Physics, Taras Shevchenko National University of Kiev,
6 Academician Glushkov Avenue, Kiev 03680, Ukraine}

\author{S.G. Sharapov}
\affiliation{Bogolyubov Institute for Theoretical Physics, National Academy of Science of Ukraine, 14-b
        Metrolohichna Street, Kiev 03680, Ukraine}

\author{V.P. Gusynin}
\affiliation{Bogolyubov Institute for Theoretical Physics, National Academy of Science of Ukraine, 14-b
        Metrolohichna Street, Kiev 03680, Ukraine}

\author{A.V. Kavokin}
\affiliation{CNR-SPIN, Viale del Politecnico 1, I-00133 Rome, Italy and Physics and Astronomy, University of Southampton, Higfield, Southampton, SO171BJ, United Kindom}

\author{A.A. Varlamov}
\affiliation{CNR-SPIN, Viale del Politecnico 1, I-00133 Rome, Italy}


\begin{abstract}
We propose a characterization tool for studies of the band structure of new materials promising for the observation of topological phase transitions. We show that a specific resonant feature in the entropy per electron dependence on the chemical potential may be considered as a fingerprint of the transition between topological and trivial insulator phases.
The entropy per electron in a honeycomb two-dimensional crystal of germanene subjected to the external electric field is obtained from the first principles calculation of the density of electronic states and the Maxwell relation. We demonstrate that, in agreement with the recent prediction of the analytical model, strong spikes in the entropy per particle dependence on the chemical potential appear at low temperatures.
They are observed at the values of the applied bias both below and above the critical value that corresponds to the transition between the topological insulator and trivial insulator phases, whereas the giant resonant feature in the vicinity of the zero chemical potential is strongly suppressed at the topological transition point, in the low temperature limit. In a wide energy range, the van Hove singularities in the electronic density of states manifest themselves as zeros in the entropy per particle dependence on the chemical potential.

\end{abstract}


\maketitle


\section{Introduction}
In recent years, a class of new topological materials has been theoretically predicted and experimentally studied (see reviews \cite{Hasan2010RMP,Qi2011RMP}).
Topological insulators are characterized by bulk band gaps and gapless edges or surface states, that are protected by the time-reversal symmetry and characterized by a $Z_2$ topological order parameter.
Novel group-IV graphenelike two-dimensional (2D) crystals such as silicene, germanene, and stanene are examples of the two-dimensional topological insulators proposed in Refs. \cite{Kane.Mele.2005,Bernevig2006PRL}.
They attract enhanced attention nowadays because of their high potential for applications in nano-electronic devices of a new generation.

In this paper, we specifically consider germanene. We propose an experimental method for the detection of an electric field induced transition between topological and trivial insulator phases of this  material.
Germanene is a two-dimensional crystal with a buckled honeycomb structure that can be considered as a germanium-based analog of graphene \cite{Acun2015JPCM}, whereas it possesses a rather large spin-orbit induced gap on the quasiparticle spectrum.
According to the recent theoretical works, germanene appears to be a natural topological $Z_2$ insulator  \cite{Liu2011PRL,Liu2011PRB}.
Yet it can be brought to the conventional (trivial) insulator phase by
applying the external electric field \cite{ohare2012stable,Drummond.2012,ezawa2012topological,matthes2014influence,ezawa2015monolayer} perpendicular to its
plane which induces a second energy gap owing to the breaking of the inversion
symmetry due to the field and buckling.

\begin{figure*}[t]
\includegraphics[width=\linewidth]{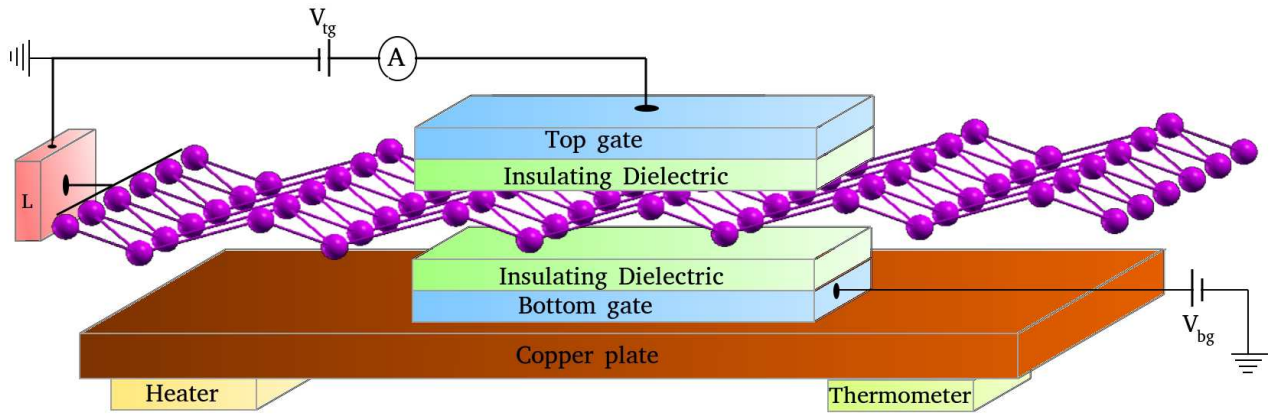}
\caption{The schematic of a possible experimental setup to measure entropy per particle, or
$\partial \mu /\partial T$,   in  germanene, or another 2D crystal.
The applied top-gate $V_{\mathrm{tg}}$  and bottom-gate $V_{\mathrm{bg}}$ voltages
allow one to control independently both the chemical potential and the perpendicular
electric field. The sample and the copper sample holder are kept in thermal contact with a
wire heater, which modulates their temperature and changes  the chemical potential of the sample.
The value $\partial \mu /\partial T$ is directly determined from
the measured recharging current between the crystal and the top electrode \cite{Pudalov2015NatComm}.
}\label{fig:setup}
\end{figure*}

In order to access the rich physics of topological transitions and to design the germanene-based nanoelectronic devices one needs precise knowledge and control over its band structure in the far infrared and terahertz spectral ranges.
The conventional experimental methods of the band structure study, in particular the optical transmission spectroscopy, may not be sufficient because of the specific optical selection rules in a gated germanene, the lack of tunable terahertz laser sources, and the low interband absorption in one-monolayer crystal structures \cite{bechstedt2012infrared,Stille2012PRB}.
Recently, a promising tool for the band structure studies with use of the electronic transport measurements has been proposed and successfully tested on two-dimensional electron gas with a parabolic dispersion \cite{Pudalov2015NatComm}.
This method, based on the measurements of recharging currents in a planar capacitor geometry, gives access to the entropy per particle $s = \partial S/ \partial n$ ($S$ is the entropy per unit volume and $n$ is the electron density).
This characterization technique is based on the Maxwell relation that links the temperature derivative of the chemical potential in the system, $\mu$, to the entropy per particle $s$:
$- \partial \mu /\partial T = \partial S/ \partial n$.

The modulation of the sample temperature changes the chemical potential and, hence, causes recharging of the gated structure where the 2D electron gas and the gate act as two plates of a capacitor.
Therefore, $\partial \mu /\partial T$ may be directly obtained in this experiment from the measured recharging current.

The entropy per particle is an important characteristic \textit{per se} of any manybody system.
The link between the measurable $s(\mu)$ and the electronic density of states (DOS) allows extracting the latter  from the numerical fit of the results of recharging current measurements with a precision that may exceed one of optical experiments in the particular case of gapless or narrow gap crystals \cite{Pudalov2015NatComm}.
The entropy per particle also governs the thermoelectric and thermomagnetic properties of the system entering explicitly to the expressions for the Seebeck and Nernst-Ettingshausen coefficients \cite{VarlamovKavokin2013,goupil2011thermodynamics}.

The entropy per particle in 2D fermionic systems is expected to exhibit a strongly non-monotonic dependence on the chemical potential in some cases.
Recently, it has been theoretically predicted \cite{Varlamov2016PRB} that, in a quasi-two-dimensional electron gas with a parabolic dispersion, the entropy per electron should exhibit quantized peaks where the chemical potential crosses the size quantized levels.
The amplitude of such peaks in the absence of scattering depends only on the subband quantization number and it is independent of the material parameters, the shape of the confining potential, the electron effective mass, and temperature.
Very recently the behavior of $s$ as a function of the chemical potential, temperature, and the gap width for gapped Dirac materials  has been studied analytically, using the model quasiparticle spectrum \cite{Tsaran2017SciRep}.
Special attention has been paid to low-buckled Dirac materials \cite{Liu2011PRL,Liu2011PRB}, e.g., silicene \cite{Kara2012SSR} and germanene \cite{Acun2015JPCM}.
It has been demonstrated  that the resonant dip and  peak structure at the zero chemical potential and the entropy spikes in the $s(\mu)$ dependence characterize any fermionic system with multiple discontinuities in the DOS.

In the present paper, we study the entropy per electron dependence on the chemical potential of the electron gas in
a  germanene crystal subjected to the external electric field applied perpendicularly to the crystal plane.

Figure \ref{fig:setup}  shows a possible experimental setup to measure the entropy per particle or
$\partial \mu/\partial T$, using dual-gated geometry for band gap engineering (see, for example, Ref. \cite{Wang2011PRL}).
The top-gate $V_{\mathrm{tg}}$  and bottom-gate $V_{\mathrm{bg}}$ voltages,  are applied to change the density of the carriers and the perpendicular electric field, independently.
Time modulation of the sample temperature changes the chemical potential and leads to the current flow between the germanene sheet and the top gate. The entropy per particle as a function of the chemical potential would be extracted from the recharging currents measurements as described in Ref. \cite{Pudalov2015NatComm}.

Our paper is organized as follows: In Sec. II we describe the theoretical approach used to model the entropy per particle and explain how we calculate the electronic properties of germanene from the first principles. In Sec. III we present the results of modeling and show how the entropy per particle dependence on the chemical potential is changing at the transition point between the topological insulator and the trivial insulator phases. Finally, the conclusions are given in Sec. IV.

\section{Theoretical  Methods}

\subsection{Entropy per particle}

The {\it ab initio} calculations of the germanene electronic band structure allow extracting the value of critical electric-field $E_c$ where the transition between the topological and the trivial phases takes place. The same {\it ab initio} calculations provide us by the detailed dependence of the one-electron DOS per spin,
\begin{equation}
	\label{eqn:DOS}
    D(\varepsilon) = \sum_{i=1}^N \int_{BZ} \frac{d{\bf k}}{(2\pi)^2} \delta \left( \varepsilon - \varepsilon_{i,{\bf k}} \right) ,
\end{equation}
with $i$ running over all the bands up to the $N$th one.
In its turn $D (\varepsilon )$ can be related to the entropy per particle.  For the Fermi system this dependence is given by a general relation \cite{Varlamov2016PRB,Tsaran2017SciRep},
\begin{equation} \label{entropy-part}
s (\mu,T)
=\frac1T\frac{\int _{-\infty}^{\infty }d \varepsilon D (\varepsilon ) (\varepsilon - \mu) \cosh^{-2}
\left( \frac{\varepsilon - \mu}{2T} \right) }{\int _{-\infty}^{\infty }d \varepsilon D (\varepsilon )   \cosh^{-2} \left( \frac{\varepsilon - \mu}{2T} \right) },
\end{equation}
where we set the Boltzmann constant $k_B=1$.

Note that the entropy per particle is rather sensitive to the electron-hole asymmetry [see numerator of Eq. (\ref{entropy-part})]
and plays an important role for the detection of the Lifshitz topological transition \cite{Blanter1994PR}. Whereas in the DOS or conductivity the Lifshitz transitions manifest themselves as weak cusps,
in the thermoelectric power [directly related to $s(\mu,T)$] giant singularities occur. Such a singularity can be considered as a smoking gun of the topological Lifshitz transition.
Intuitively, one can expect that the sensitivity of $s(\mu,T)$  to topological transformations of the Fermi surface would give an opportunity to trace out various types of topological transitions. In the present paper we demonstrate that $s(\mu,T)$ indeed offers the opportunity to identify the transition between the topological and the trivial insulator, whereas the entropy feature characterizing this transition is quite different from one characteristic of the Lifshitz transition.

\subsection{{\it Ab initio} calculations of the DOS }

 Our calculations of the DOS and the electronic band structure of germanene are based on the density functional theory (DFT) as implemented in the QUANTUM ESPRESSO package \cite{espresso.2009,espresso.2017}.
The single-particle Schr\"{o}dinger equation as formulated by Kohn-Sham\cite{kohn.sham.1965}
 \begin{equation}
\label{eqn:Kohn-Sham}
\begin{split}
\left(\! -\!\frac{\hbar^2}{2m}\nabla^2\! +\! v_{\mathrm{ext}}({\bf r})\! +\! \int \! \frac{n({\bf r'})}{\left| {\bf r} \!- \!{\bf r'} \right|} d{\bf r'} \!+\! v_{\mathrm{xc}}({\bf r})\!\right)\psi^{\mathrm{K\!S}}_{i,{\bf k}}\!=\!\varepsilon^{\mathrm{K\!S}}_{i,{\bf k}} \psi^{\mathrm{K\!S}}_{i,{\bf k}}
\end{split}
\end{equation}
($v_{\mathrm{ext}}$ is the electron-ion potential and $v_{\mathrm{xc}}$ is the exchange-correlation potential) is solved self-consistently through the wave-functions' expansion on plane-wave basis sets with use of the periodic boundary conditions.
  For the germanium atoms we use the norm-conserving scheme\cite{hamann.schluter.1979}, the valence electronic configuration $3d^{10} 4s^2 4p^2$, and the generalized gradient approximation Perdew-Burke-Ernzerhof\cite{PBE.1996}(GGA-PBE) for the exchange and correlation potential.
  After accurate convergence tests on the total energy results,  an energy cut-off of $90\, Ry$ has been selected.

We model our 2D crystal  as an infinite $xy$ plane of germanium atoms in the honeycomb geometry.  The theoretical lattice constant for the hexagonal cell, obtained by minimization of the total energy, was determined to be $a=4.04 \, \angstrom$ with a low buckling configuration ($\delta_{LB} = 0.68 \, \angstrom$), in agreement with previous results\cite{prl.ciraci.2009}. Since the periodic boundary conditions are being enforced over all axes, the use of  supercells large enough  to avoid spurious interactions between periodic images is required. After tests over the computed potential and energies, we use a supercell containing $32 \, \angstrom$ of vacuum along the $z$ direction.

 In the absence of the spin-orbit (SO) interaction the germanene spectrum represents a perfect Dirac cone characterized by gapless fermions  with the Fermi velocity of about $0.5 \times 10^6 \, m/s$.
 By switching on the SO interaction a small gap $\Delta_{SO}$ opens at the $K$, $K^\prime$ points of the Brillouin zone (BZ), and the bands linearity is lost, leading to the appearance of gapped fermions\cite{SO,Corso.Conte.2005,Conte.Fabris.2008}.
 We obtain a value of the gap of $24\,meV$, in agreement with the previous GGA-PBE results. It is slightly below the value of $33 \, meV$ found with use of non-local hybrid exchange and correlation functionals \cite{matthes2013massive}.

 The gaps in the electronic spectrum can be further modified by applying an external field (bias) perpendicular to the germanene plane.
 This is performed by superimposing a sawtooth potential along the $z$ direction of the crystal.
 The properties of the system have been studied for different values of the applied bias, ranging from $0$ to $0.4 \,V/\angstrom$.
For the DOS calculations a very high energy resolution is needed in order to observe the small differences in low energy features induced by the different electric fields.
For this reason, we used  a refined mesh of $12000 \times 12000 \times 1$  Monkhorst-Pack \cite{Monkhorst.Pack.1976} $k$-points in the BZ cropped around $K$($K^\prime$) with a radius of $0.02 \times 2\pi/a$.

A topological phase transition should be observed at the specific value of the applied field $E=E_c$, where the fundamental electronic gap closes.
As was mentioned before, at the electric fields below this value, germanene is a topological insulator whereas above the critical field it becomes a trivial insulator \cite{ezawa2012topological,matthes2014influence}.
In order to prove it, we calculate the topological invariant $Z_2$ [\onlinecite{fu.kane.2006}]. This invariant is indicative of the realization of the quantum spin Hall effect\cite{Kane.Mele.2005}. It acquires the values of 0, for a trivial phase, or 1 for a topological phase.
In a system that obeys the inversion symmetry, $Z_2$ can be defined as\cite{fu.kane.2007}:
\begin{equation}
	\label{z2}
	(-1)^{Z_2} = \prod_{n=1}^N \xi_{2n} \left(K\right),
\end{equation}
where $\xi_{2n} \left( K \right)= \pm 1,$ is the eigenvalue of the parity operator calculated at the time reversal invariant momenta(TRIM) points for each $2nth$ band. One easily sees that $Z_2$ can acquire only two values: $Z_2 = 0,1$.

The topological invariant $Z_2$ has been computed here by applying the Wilson loop method, following the evolution of the Wannier Charge Centers in a plane containing the TRIM points\cite{Kane.Mele.2005,Soluyanov.Vanderbilt.2011,wilson.loop.2011}, as implemented in Z2PACK\cite{z2pack.2017}.
This method yields a correct result even if the inversion symmetry is broken by the application of an electric field.

\section{Results and Discussion}

\subsection{Effect of spin-orbit and Electric field on the electronic properties}
In the absence of external electric fields, germanene crystals possess the inversion symmetry which, together with time reversal symmetry, leads to the spin degenerate energy bands.
The application of bias brakes the inversion symmetry causing the lifting of the spin degeneracy.
The eigenvalues for each band are computed by solving the Kohn-Sham equation \eqref{eqn:Kohn-Sham}. The corresponding band structure at K and K$^\prime$ points can be approximated by the relation \cite{Kane.Mele.2005,Drummond.2012}
\begin{equation}
\label{eqn:band_field}
\varepsilon_{\eta,\sigma}^{c/v} \ = \ \pm\frac{1}{2}(\Delta_{SO} \ - \sigma\eta\Delta_{el}).
\end{equation}
Here $\Delta_{SO}$ is the spin-orbit splitting of $24 \, meV$, $\Delta_{el}$ is the splitting induced by the electric field, $\eta = \pm 1$ and $\sigma = \pm 1$ are the valley ($K$ and $K^\prime$) and spin ($\uparrow, \downarrow$) subscripts, respectively, and  $c$ and $v$ denote the conduction and valence bands corresponding to the $\pm$ signs in the equation.

 As is seen from Eq. \eqref{eqn:band_field} four linear bands show up now at the $K$, and four at $K^\prime$ points. They are separated in energy as described by a spin-dependent gap,
\begin{equation}
\label{eqn:spin_gap}
\Delta _{\eta,\sigma }= \varepsilon_{\eta,\sigma}^{c} - \varepsilon_{\eta,\sigma}^{v}.
\end{equation}
Since the time reversal symmetry is still present,  the relation $\varepsilon_{\eta,\sigma} = \varepsilon_{-\eta,-\sigma}$ for the band dispersion holds.

\begin{figure}[!ht]
  \includegraphics[width=\linewidth]{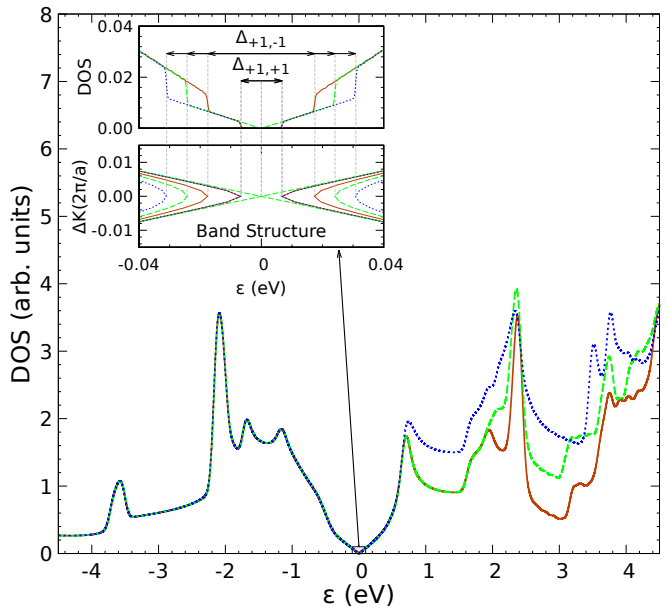}
  \caption{ Density of states computed within the DFT for the external electric fields below/at/above the critical value $E_c$:  $E = 0.10 \,V/\angstrom$ (orange solid line), $0.23 \,V/\angstrom$ (green dashed line) and $0.36 \,V/\angstrom$ (blue dotted line). The zero denotes the Fermi energy.
  The insets: zoom of the DOS near the Fermi energy (upper panel) and the electronic band structure in a close proximity of the high symmetry point K (lower panel).
}\label{fig:DOS+zoom}
\end{figure}

The DOS is   calculated using Eq. \eqref{eqn:DOS}. The corresponding results are shown in Fig. \ref{fig:DOS+zoom} for a wide energy range. A broadening of the $\delta$ function in Eq. \eqref{eqn:DOS} of $50 \,meV$($0.3 \,meV$ for the upper inset) has been used.
We can observe that the low-energy DOS exhibits the expected linear behavior for the unbiased germanene. The applied electric field leads to the appearance of the gaplike feature in the DOS.

\begin{figure*}[t]
  \includegraphics[width=\linewidth]{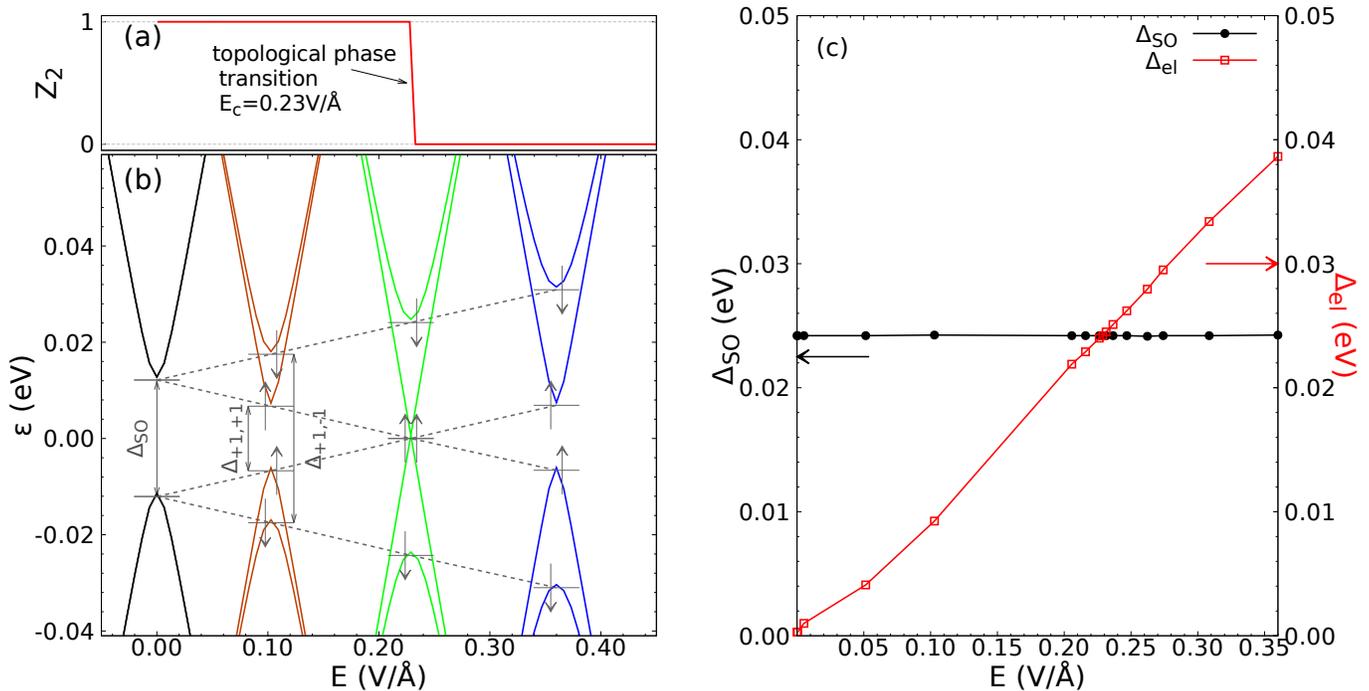}
  \caption{
  (a) Computed topological invariant ($Z_2$) at different values of the external electric field.
  (b) The electronic band structure near the $K$-point of the Brillouin zone computed at different values of the external electric field.
  (c) The variation of the spin-orbit ($\Delta_{SO}$) and electric-field($\Delta _{el}$) induced splittings with the increase in the external electric field.}\label{fig:Gaps2}
\end{figure*}

By using Eq. \eqref{eqn:band_field} one can obtain the values of $\Delta_{SO}$ and $\Delta_{el}$ for different electric fields as shown in Fig. \ref{fig:Gaps2}(c).
It can be seen that the former is independent of the field $E$ and remains equal to $24\,meV$, whereas the latter shows a linear dependence on $E$. The splitting $\Delta_{el}$ reaches the value of  $\Delta_{SO}$ at $E_c=0.23\, V/\angstrom$ which marks the critical field for the topological phase transition.

We can see in Fig.~\ref{fig:Gaps2}~(c) that the electronic gap $\Delta_{el}$ depends linearly on the applied electric field.
For the values of the field below the critical one ($E < E_c$) the smallest (fundamental) gap decreases, until it closes up completely at $E = E_c$.
In this regime, germanene is a topological insulator, whereas, for larger field values ($E > E_c$), the smallest  electronic gap opens up again,
and germanene  becomes a trivial insulator.

In what concerns the topological state of germanene we found that $Z_2=1$  for all values of electric field below $E_c$ and $Z_2=0$ otherwise, as
shown in Fig.~\ref{fig:Gaps2}a.


\subsection{Dependence of the entropy on the applied electric field}

As has been shown in the previous work \cite{Tsaran2017SciRep}, in the case of a crystal characterized by two nonzero energy gaps,
the dependence $s(\mu)$ exhibits two distinct structures in both the electron, $\mu >0$, and the hole, $\mu <0$, doped regions.
The first one is a giant resonant feature in the vicinity of the zero chemical potential, and the second one is a spike of the height
$s =  2 \ln 2/3$ at the edge of the larger gap. These resonances are apparent in the low-temperature limit.

\begin{figure}[!ht]
\includegraphics[width=\linewidth]{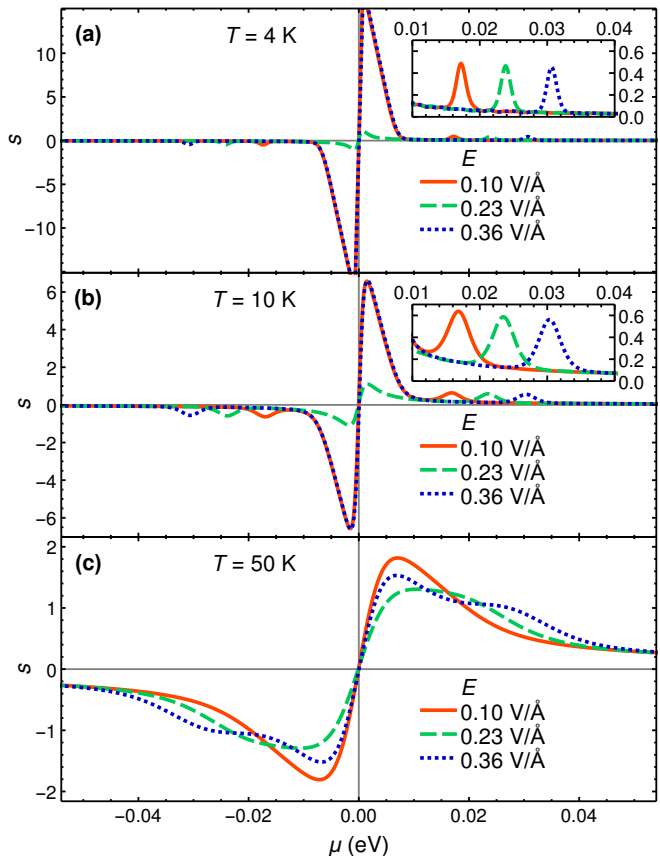}
\caption{The entropy per electron $s$ vs the chemical potential $\mu$ in eV in the vicinity of the
Dirac point for three values of electric field $E$:
(a) $T=4 \mbox{K}$ (b) $T = 10 \mbox{K}$ (c) $T = 50 \mbox{K}$.
The insets in (a) and (b) show the zoomed domains with the entropy spikes of  the height $s = 2 \ln2/3$ at low temperatures.}
\label{fig:Spikes}
\end{figure}

Our results for the entropy per particle $s$ for germanene are presented in Fig.~\ref{fig:Spikes}.
The entropy spikes occur both above and below the transition between the topological and the trivial insulator phases as well as exactly at the transition point, $E_{c} = 0.23 \, V/ \AA$
[see the insets of Figs.~\ref{fig:Spikes}~(a), (b)].
The most prominent result of the present paper is that the strong resonant feature of the entropy per particle in the close vicinity
of the Dirac point $\mu =0$ is nearly fully suppressed at the transition point, $E_c$,  whereas it occurs for values of the
electric field below and above it.

It is important to note that the appearance of the second step in the electronic density of states due to the lifted spin
degeneracy of the  Brillouin zone has a dramatic effect on $s$.
In particular, the resonant feature in the vicinity of the zero chemical potential is strongly pronounced
if the DOS exhibits two steps (see the upper inset in Fig.~\ref{fig:DOS+zoom}). Indeed, for
$|\mu| \ll T \ll \Delta_{+1,+1}$  it was obtained \cite{Tsaran2017SciRep} that
\begin{equation}
s(T, \mu ,\Delta_{+1,+1} ) \simeq \frac{\mu \Delta_{+1,+1}}{2T^2} .
\end{equation}
For the critical field $E=E_c$ the gap $\Delta_{+1,+1} = 0$, so that the DOS exhibits
only one step and
\begin{equation}
s(T,\mu, 0 )= \frac{\mu}{T}, \quad   |\mu| \ll T.
\end{equation}
Clearly at very low temperature the peak at finite $\Delta_{+1,+1}$ is much stronger than the one for  $\Delta_{+1,+1}=0$.

The disappearance of the characteristic entropy resonance can be considered as a signature of the topological phase transition in germanene.
We are confident that this analysis would help extracting the important band parameters of topological 2D crystals from the recharging current measurements.
We also mention   that the plasmon modes and Friedel
oscillation can be used to detect
the topological phase transition in silicene and germanene even when the Fermi
level does not lie in the band
gap, as suggested by Chang $et al.$ 
\cite{Chang2014}.

\subsection{Van Hove singularities in the entropy per particle dependence on the chemical potential on a large energy scale}

\begin{figure*}[t]
\includegraphics[width=\linewidth]{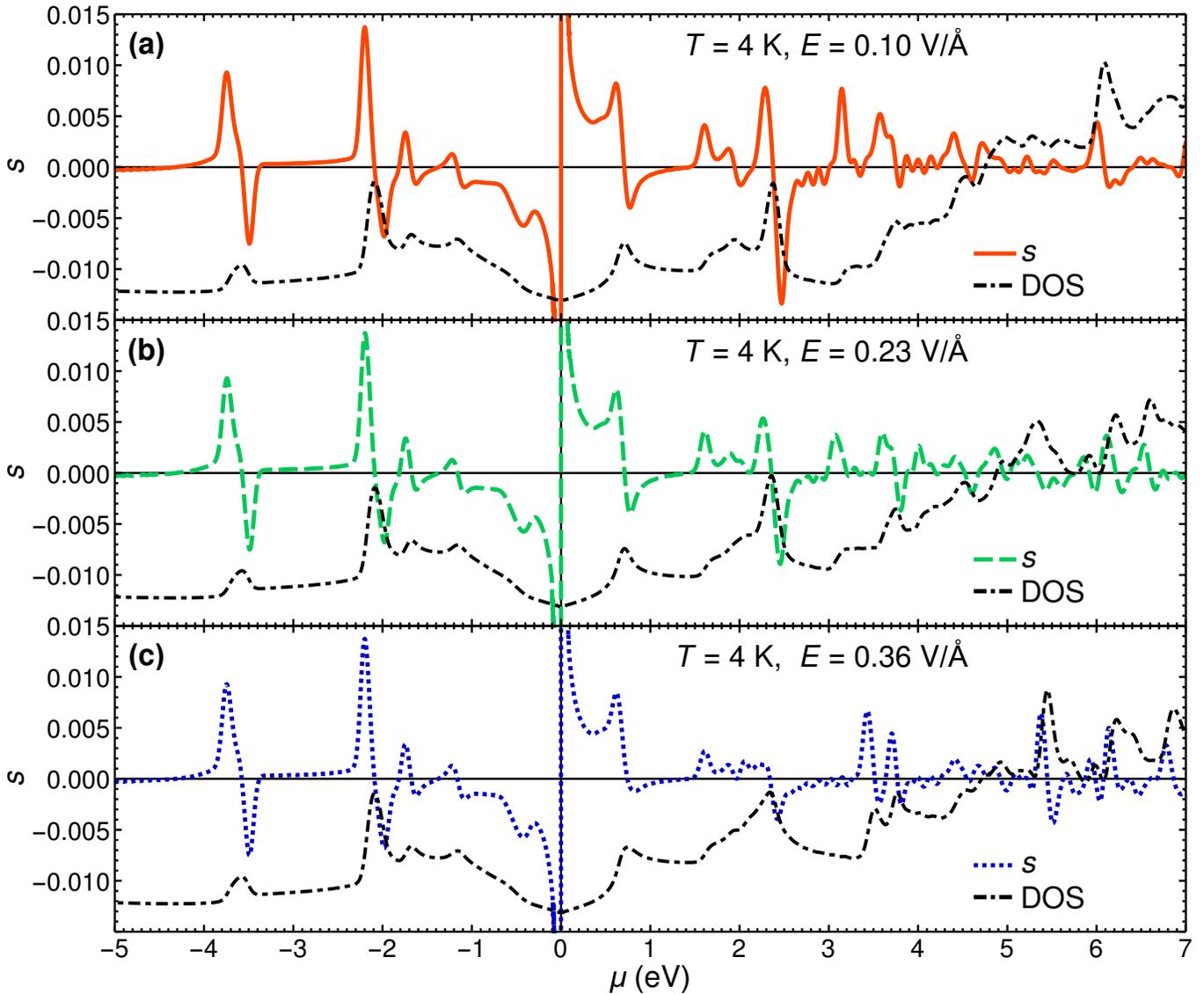}
\caption{(Color online) The entropy per electron $s$ vs the chemical potential $\mu$ in eV for three values of
electric field $E$ for $T= 4 \mbox{K}$:
(a) $E= 0.10\, \mbox{V/\AA}$ (b) $E= 0.23\, \mbox{V/\AA}$ (c) $E= 0.36 \mbox{V/\AA}$. The dashed lines show the
corresponding DOS obtained by the $ab-initio$ calculation. }\label{fig:Spikes-fullrange}
\end{figure*}

Scanning of the chemical potential on a large energy scale is challenging from the experimental view.
Possibly this could be performed by combining electrostatic and chemical dopings.
Nevertheless, the behavior of the entropy per particle at large values of the chemical potential is worth analyzing theoretically as it offers some non-trivial features. Figure \ref{fig:Spikes-fullrange} shows $s(\mu)$ in comparison with the DOS for three values of the applied field.
We note that both energy and entropy scales in Fig. \ref{fig:Spikes-fullrange} are orders of magnitude different from those
in  Fig. \ref{fig:Spikes}, so that the high-energy features in $s(\mu)$ are several orders of magnitude weaker than the resonant feature
at the zero chemical potential discussed above. In particular,
the disappearance  of the resonant feature at the critical field $E_c$  cannot be recognized in Fig.~\ref{fig:Spikes-fullrange}~(b)
due to the small range of shown values of $s$.
Nevertheless, these features represent a significant interest. Using
Eq.~(\ref{entropy-part}) it is easy to see that the extrema of the dependence of $D(\mu)$ are converted to the zeros of $s(\mu)$.
Indeed, assuming that $D(\mu)$ is a smooth function one may expand it near the extremum of $\mu_{\mathrm{ex}}$ and
obtain that $s(\mu) \varpropto D^\prime(\mu_{\mathrm{ex}})$ \cite{Tsaran2017SciRep} where the derivative
$D^\prime(\mu_{\mathrm{ex}})$ changes sign at the extrema of $\mu_{\mathrm{ex}}$.
Since van Hove singularities correspond to the sharp peaks in the DOS, they show up in the dependence
$s(\mu)$ as the strong positive peak  and negative dip structures.
Note that even the discussed above the giant negative dip and positive peak structure near the zero chemical potential might be interpreted as a reflection of the negative $V$- and $U$-like shape peaks of the DOS.
This demonstrates that the experimental investigation of $s(\mu)$ in a wider range of energies can be useful for
tracing out van Hove singularities and their evolution in the perpendicularly applied electric field.

\section{Conclusions}

We have studied fingerprints of the topological phase transitions
and DOS singularities in two-dimensional materials with use of the $ab-initio$ calculations. We show that the entropy per particle dependence on the chemical potential is highly sensitive to the DOS. 
In particular, at the critical field corresponding to the transition point between topological and trivial insulator phases, the strong resonant feature of the entropy per particle at the zero chemical potential disappears. Moreover, at the Van Hove singularities of the DOS the entropy per particle passes through zero (dip-peak features). Based on these theoretical findings, we propose an experimental method of detection of the critical transition points and density of states singularities in novel structures and materials.

The found characteristic feature of the entropy per particle close to the transition between the trivial and the topological insulators might become the smoking gun also for other type of topological transitions in novel systems. 
For instance, in the recent work of Wang and Fu  \cite{wang2017topological} the authors demonstrated the possibility of  transition between the trivial and the topological superconductors \cite{sato2017topological}. 
It is worth noting that in superconductors with different pairing symmetries placed in a magnetic field the vortex entropy is different. Being subjected to the gradient of temperature, the vortices move along the gradient, resulting in the appearance of the flow of
entropy, which can be directly measurable by the Nernst signal. 
Hence, one could expect that the latter can be an appropriate tool for the study of such
transitions.

\section*{Acknowledgements}
We acknowledge support from the HORIZON 2020 RISE "CoExAN" Project (Project No. GA644076).
S.G.Sh. and V.P.G. acknowledge a partial support from the National Academy of Sciences of Ukraine (Projects No. 0117U000236 and No. 0116U003191) and by its Program of Fundamental Research of the Department of Physics and Astronomy (Project No. 0117U000240).


\begin{thebibliography}{37}
\bibitem [{SO()}]{SO}%
  \BibitemOpen
  \href@noop {} {\enquote {\bibinfo {title} {{Spin Orbit (SO) interaction has
  been included by, first, solving the radial Dirac equation for each isolated
  atom and, second, by reducing the four-component Dirac spinors to
  two-component Pauli spinors in order to generate pseudopotentials with
  two-component projectors}},}\ }\BibitemShut {NoStop}%
\end{thebibliography}
\end{document}